# Precise magnetization measurements down to 500 mK using a miniature $^3$He cryostat and a closed-cycle $^3$He gas handling system installed in a SQUID magnetometer without continuous-cooling functionality

Kazutoshi Shimamura[1,2]*, Hiroki Wajima[1], Hayato Makino[1], Satoshi Abe[1], Yoshinori Haga[3], Yoshiaki Sato[4], Tatsuya Kawae[4] and Yasuo Yoshida[1]*

[1]*Department of Physics, Kanazawa University, Kanazawa, Ishikawa 920-1192, Japan*

[2]*Instrumental Analysis Division, Engineering and Technology Department, Kanazawa, Ishikawa 920-1192, Japan*

[3]*Advance Science Research center, Japan Atomic Energy Agency, Tokai, Ibaraki, 319-1195, Japan*

[4]*Department of Applied Quantum Physics, Faculty of Engineering, Kyushu University, Fukuoka 819-0395, Japan*

E-mail: shimamura@se.kanazawa-u.ac.jp, yyoshida@se.kanazawa-u.ac.jp

We have conducted precise magnetization measurements down to 0.5 K with a miniature $^3$He cryostat and a closed-cycle $^3$He gas handling system for a commercial superconducting quantum interference device magnetometer [Magnetic Property Measurement System (Quantum Design)]. The gas handling system contains two sorption pumps filled with granular charcoals. We pressurize $^3$He gas up to ambient pressure for the liquification at 3 K and then pump the vaper for cooling. The lowest sample temperature is ~ 0.5 K and it can persist for 34 hours. We demonstrate the performance of the system by observing the Meissner effect of aluminum below the superconducting transition temperature ~ 1 K. We also measured the magnetization curve of the heavy fermion superconductor CeCoIn$_5$ resulting in successful observation of the lower critical field at 0.5 K.

## 1. Introduction

Magnetization measurement at low temperatures is one of the fundamental methods to explore the intrinsic magnetic and superconducting properties of materials (e.g. Pd[1], Au[2], Pt[3], Cr[4]). Among commercially available magnetometers, the magnetometer called Magnetic Property Measurement System (MPMS) has been widely used in various research fields. Thanks to the high sensitivity of the superconducting quantum interference device (SQUID), MPMS can detect small magnetic moments in the order of $10^{-7}$ emu. Another advantage of MPMS is the capability of measurements in the wide magnetic field and temperature ranges; 0 Oe $\leq H \leq$ 70 kOe, 1.8 K$\leq T \leq$ 400 K. In addition, other external parameters like high pressures and external electric fields are applied to samples during magnetization measurements in order to tune the physical property of materials.[5-12] It has been also utilized to estimate the penetration depth of a ring-shaped superconducting sample whose magnetization is induced by the vector potential around a long miniature solenoid coil threaded through the sample[13].

Although various efforts have been paid for magnetization measurements in various conditions, measurements below 1.8 K have not been performed before the pioneer works by iQUANTUM Corp.[14),15)] and Sato *et al* [16]. They developed miniature $^3$He cryostats directly installable to MPMS. $^3$He gas is liquefied in the cryostats by setting the MPMS temperature to 2 K, and then the cryostat is cooled down by pumping liquid $^3$He with an external pump. In the case of iQuantum, a sample moves with the whole metallic cryostat for measurements, which limits the resolution of the magnetometry to $10^{-4}$ emu when an external field is more than 1 T although the background can be subtracted.[17] Sato *et al*. have the better design in which only a sample moves to hold the high resolution below $\sim 10^{-7}$ emu.[16] Those authors use MPMS systems which can keep 2 K during the whole process of experiments by pumping the $^4$He liquid reservoir where $^4$He liquid continuously come in from the liquid $^4$He bath. However, several types of MPMS have been on market so far and the old models which cannot maintain 2 K for long time are still widely used. These old models do not have the continuous cooling functionality, and pump on the liquid $^4$He reservoir only after it is filled. Once the reservoir gets empty the MPMS temperature increases up to more than 7 K. A typical duration for which this type of MPMS can maintain 2 K is only about 3 hours. Therefore, it is challenging to perform measurements at lower temperatures even with the well-designed miniature cryostat mentioned above. We overcome this difficulty by liquefying $^3$He gas at higher temperature $T =$ 3 K and at higher pressures close to the ambient pressure. A high-pressure compatible gas handling system (GHS), originally made for $^3$He melting curve thermometry, enables us to produce high pressure $^3$He gas. In this paper, we report the designs of the cryostat and GHS, the results of test cooling, and the measurement results of aluminum and a heavy fermion superconductor.

## 2. Experimental methods

### 2.1 Design of miniature $^3$He cryostat

Figures 1(a) and 1(b) show the design of our miniature $^3$He cryostat. The main part of the cryostat is made with a stainless-steel pipe with 6.5-inch outer diameter. It goes from the room temperature part all the way down to the $^3$He pot at low temperatures. $^3$He gas is introduced into and pumped out from the sample and $^3$He line indicated with purple marker in Fig.1(a). In the low-temperature area, copper and stainless-steel pipes are brazed with a silver solder to the copper-made top flange (Fig. 1(b)) of the inner vacuum chamber (IVC). A stainless-steel capillary is soldered also to the copper top flange for the IVC pumping line as indicated with olive marker in Fig.1(a). The IVC is for thermally isolation of the bottom of the stainless-steel pipe from the low-temperature area of MPMS. The stainless-steel pipe inside the IVC is cooled via the top flange and $^3$He gas is condensed in the area surrounded by a Cu foil, which works as the $^3$He pot (Fig.1(b)). We use the Cu foil to cool down liquid $^3$He homogeneously and not to have temperature gradients due to the bad thermal conductivity of liquid $^3$He[18]. The stainless-steel pipe is capped with a copper cap at the bottom where a RuO$_2$ resistor is attached. The thermometer leads are thermally anchored to the stainless-steel pipe inside the IVC (Fig.1(b)).

### 2.2 Design of Gas Handling System (GHS)

Figure 2(a) is the sketch of our GHS and a miniature $^3$He cryostat[16]. A 1/16-inch stainless steel tube is mainly used for the GHS to restrict the available volume for $^3$He gas. The GHS stores 3 litter of $^3$He gas in the gas storage and equips two sorption pumps made of a stainless steel pipe with 1/2-inch in diameter filled with charcoals. The sorption pumps 1 and 2 are connected to the GHS with a 1/16-inch stainless-steel tube and a 1/2-inch flexible tube, respectively. The diameter of the sorption pumps fit to the inlet diameter of liquid $^4$He storage vessels and therefore we can easily cool them at 4 K by dipping them in liquid $^4$He. We use the sorption pumps both for pumping and pressurizing $^3$He gas. Sorption pumps utilize molecular adsorption at the activated charcoal with the large surface area. $^3$He gas is adsorbed at the charcoal when the sorption pump is cooled down to 4 K. By warming up the pumps to room temperatures after adsorbing $^3$He gas, we can pressurize $^3$He gas up to high pressures if the available volume for $^3$He gas is restricted. Another advantage of the GHS with the sorption pumps is that there is no worry about losing precious $^3$He gas, which is always a concern when using mechanical pumps for cooling. A liquid nitrogen (LN$_2$) trap is equipped to remove other molecules (O$_2$ and N$_2$) contained in $^3$He gas. We normally drive $^3$He gas through the LN$_2$ trap to remove other molecules in $^3$He gas in the beginning and at the end of experiments. The LN$_2$ trap can be bypassed to reduce the available volume for $^3$He gas inside the GHS. Actual setting of the whole system is presented in Fig. 2(b).

## 2.3 Recording the $^3$He pot temperature

A RuO$_2$ resistor attached to the bottom of the $^3$He pot was calibrated against a calibrated Cernox thermometer attached on a silver plate at the sample rod. We monitor the resistance of the RuO$_2$ with a standard temperature controller and record in our LabVIEW program installed to the MPMS operating computer. During magnetization measurements, we record the $^3$He pot temperature every few seconds. By comparing the timestamps of the magnetization and the temperature data, both data are easily associated to each other.

## 2.4 Preparation before and after installing the $^3$He cryostat into MPMS

Before installing the $^3$He cryostat into the MPMS sample space, the IVC are evacuated. The $^3$He pot is also evacuated after introducing a sample into the $^3$He pot with a sample rod. The cryostat is installed to the MPMS sample space with the MPMS temperature set to 300 K followed by evacuating the area between the cryostat and the GHS with an external mechanical pump. The details of the cryostat design and set-up procedures are described in Ref. 16.

## 2.5 Standard procedures of condensation and cooling

To introduce high-pressure $^3$He gas to the cryostat, $^3$He gas in the gas storage first has to be pumped with the sorption pump 1. It takes about one night to complete this step. We normally use this time to precool the cryostat to 3 K by introducing small amount of $^3$He gas to the cryostat as exchange gas and set the MPMS temperature to 3 K. After all gas is pumped by the sorption pump1 and the cryostat is well cooled down to 3 K, we close the valve to the gas storage and bypass the LN$_2$ trap in order to reduce the available volume for $^3$He gas. By warming up the pump to room temperatures, we can transfer high pressure gas to the cryostat.

Condensation of ³He gas occurs based on the pressure-temperature phase diagram[18,19]. The critical temperature at 1 bar is 3.2 K and therefore we can liquefy ³He gas by introducing ³He gas at ~ 1 bar to the cryostat maintained at 3 K. Not only that, but we can also liquefy ³He gas more efficiently at 3 K than at 2 K. This is because the cooling power of ⁴He evaporation is larger owing to larger latent heat and higher vapor pressure at 3 K than those at 2 K[18]. We can also make use of the large enthalpy of cold ³He gas to cool down well the low-temperature parts of the cryostat. This leads considerable reduction of residual heat which often remains at the stainless-steel pipe in the IVC and ³He-pot thermometer leads whose thermal conductivities are bad. If we cool down the cryostat at 2 K without pressurizing ³He gas, the efficiency of the liquification is lower and residual heats cannot be removed well, resulting in drying up liquid ³He when the MPMS temperature rises after ~ 3 hours.

After the condensation is completed, we can start the ³He cooling process using the sorption pump 2. The lowest temperature ~ 0.5 K is easily obtained in a few hours because of the high conductance of the pump (the short distance from the cryostat and the wider diameter of the intermediate tubes). After the initial condensation and cooling with sorption pumps 1 and 2, we can normally carry out the condensation at 2 K with the sorption pump 2 without pressuring ³He gas. The ³He pot temperature will not rise easily even with the slow introduction of room-temperature ³He gas once ³He gas is well liquefied in the ³He pot. This is probably due to considerable reduction of residual heat in the low-temperature area through the initial condensation. The large specific heat of the liquid ³He, which come from the small fermi energy $E_F$ of ³He, $E_F/k_B$ ~ 1K, and the specific heat of a Fermi liquid $C \propto \frac{T}{T_F}$, also plays a significant role for this[18]. At the end of experiments, ³He gas is pumped with the sorption pump 1 passing through the LN₂ trap, and then stored in the gas storage.

## 3. Results and discussion

### 3.1 Cooling test

・ Initial condensation and cooling with sorption pumps 1 and 2

In figures 3(a) and (b), we present the ³He pot temperature and GHS pressure variations with time as a typical example of initial cooling of the cryostat using two sorption pumps. Before starting up experiment, we kept the MPMS and the ³He pot temperatures at 3 K. The whole cooling process of the cryostat can be divided into three steps, pressurizing, condensation, and cooling of ³He gas as shown with different colors in Figs. 3(a) and (b). By warming up the sorption pump 1 (indicated by a solid arrow at $t = 0$), the pressure of the GHS quickly increased as can be seen in Fig. 3(b). The cryostat temperature also rose up to ~ 3.7 K due to flow of room temperature $^3$He gas into the cryostat. After a while, the pressure of GHS stabilized and the cryostat temperature came back to 3 K as indicated by a solid arrow at $t =$ 0.5 hour. As condensation progresses, the pressure gradually decreases. To keep constant $^3$He gas pressure while supplying the gas into the cryostat, we warmed up the sorption pump 1 gradually. To have enough liquid $^3$He for cooling, we decreased the MPMS temperature gradually from 3 K to 2.6 K, which also lowered the $^3$He pot temperature as shown by a broken arrow in Fig 3(a). Once the pressure stabilized at the end of the condensation indicated by a broken arrow in Fig. 3(b), we started decreasing the pressure of the GHS by dipping the sorption pump 2 into the liquid $^4$He vessel as indicated by a solid arrow at $t = 3$ hours, and eventually started pumping the $^3$He pot for cooling. The time delay between drops of the pressure and the temperature merely comes from the delay of valve openings between the sorption pump 2 and the cryostat. It took only about 10 mins for the $^3$He pot temperature to reach ~ 0.6 K and did another 40 minutes to reach the lowest temperature ~0.5 K.

・ Condensation and cooling from the second time only with sorption 2

Figures 4(a) and (b) show a typical time dependent variation of the MPMS and the $^3$He pot temperatures in the condensation process at 2 K by using only the sorption pump 2 after the initial condensation and cooling. From the second time, we can condense $^3$He gas without ~ 1 bar pressuring because of the large specific heat of liquid $^3$He and considerable reduction of residual heats in the low-temperature area at the initial process as mentioned above. It takes about 10 hours for full condensation. During the condensation, the MPMS temperature periodically increases due to drying up of the liquid $^4$He reservoir. However, the $^3$He pot temperature is little affected by these events. As shown in the inset of Fig. 4(b), while the MPMS temperature rises from 2 K to 5.5 K and it takes about 40 minutes to return to the set point temperature ($T = 2$ K), the $^3$He pot temperature remains ~ 2 K. The large specific heat of liquid $^3$He at low temperatures prevents the temperature from rising even with thermal perturbations. This allows us to condensate $^3$He much easier since we can maintain the $^3$He pot temperature at ~ 2 K for long time.

Figures 4 (c) and (d) show typical temperature variations of the MPMS and $^3$He pot with time in the $^3$He cooling process. The $^3$He pot temperature remains below ~ 0.6 K for 34 hours (Fig. 4(d)) while the MPMS temperature increases from 1.8 K to ~5.1 K (at most ~ 7 K) every ~ 3 hours (Fig. 4(c)). The MPMS takes about an hour to recover the set point temperature T ~ 1.8 K once the temperature increased as shown in the inset of Fig. 4(c). Despite the changes in the MPMS temperature, the variation of the $^3$He pot temperature weakly takes place. As in the inset of Fig. 4(d), the $^3$He pot temperature falls within the range from 0.5 K to 0.6 K even when the MPMS temperature sometimes exceeds 7 K.

3.2 Observation of superconducting transition of Al wire

We performed magnetization measurements of aluminum (Al) to observe the Meissner effect associated with the superconducting transition at ~ 1.2 K. We used 6.4 mg of Al wire (99.99 % purity). The sample was fixed to a straw with 3.5-mm diameter. We first applied magnetic field 5 Oe to the sample at high temperatures and measured the magnetization continuously while cooling down the cryostat from 1.8 K to the lowest temperature ~ 0.6 K. To decrease the $^3$He pot temperature slowly, it is important to control the evacuation rate of the sorption pump 2. In the GHS, the evacuation rate can be controlled by carefully adjusting the needle valve in front of the pump. Once we find the proper evacuation rate, we barely need to change the setting, which means that a measurement is semi-automated. Figure 5(a) shows the temperature dependence of magnetization of Al. The magnetization suddenly drops showing a diamagnetic signal below $T$=1.1 K indicating the occurrence of perfect diamagnetism due to the superconducting transition. The onset is at the slightly lower temperature than the reported critical temperature of Al, 1.2 K. According to the previous works[20,21], Al shows the superconducting transition at 1.1 K in the magnetic field of 10 Oe. This field value is roughly match to the typical remanent field value of MPMS superconducting magnets[22]. Therefore, the observed lower critical temperature is most likely due to the remnant magnetic fields of the superconducting magnet.

The time dependent variations of the $^3$He-pot and MPMS temperatures and the magnetization during the measurement are shown in Figs. 5 (b, c, d). The measurement duration from 1.8 K to the lowest temperature is only 1.2 hours.

3.2 Magnetization curve of CeCoIn$_5$ single crystal

A type 2 superconductor has the lower and upper critical fields. The lower critical field ($H_{C1}$) is defined as the magnetic field where the field starts penetrating superconductors as vortices. Generally, $H_{C1}$ is less than a few hundred Oe. Since MPMS can apply a magnetic field with a 0.1 Oe interval from 0 to 70 kOe, $H_{C1}$ can be estimated by measuring magnetization curve at low enough temperatures compared to the superconducting critical temperatures. Further, superconducting volume fractions can be estimated for superconductors with the critical temperatures below 1.8 K. This is another advantage of lowering measurement temperature of MPMS below 1.8 K. The faraday method, which is the high precision magnetometry compatible to the very low temperatures and high magnetic fields[23], requires the field gradient ~ 10 T/m to detect magnetizations, which limits the minimum field at the sample position to ~ 1000 Oe and makes it difficult to estimate $H_{C1}$ with this method. We performed field-dependent magnetization measurements of the unconventional superconductor $CeCoIn_5$ [24,25] at 0.53 K. A planar single crystal (0.6×0.4×0.1 $mm^3$) whose weight is 1.22×$10^{-3}$ g was used with a straw with a 3.5 mm-diameter. An external magnetic field was applied parallel to the ab plane of the sample. For the data analysis, the demagnetization correction was performed using the ellipsoidal model.

Figure 6 (a) shows the combined graph of the magnetization and the $^3$He pot temperature data as a function of time during magnetization measurement. As shown in the previous sections, the MPMS temperature periodically increases within the range from 2 to 7 K, which affects the periodic temperature rises of the $^3$He pot from 0.53 K to 0.8 K appearing as spikes of the temperature data. To keep the same measurement temperature at each field, we add a waiting time in the measurement sequence so that the MPMS will pause the measurement in such cases until the MPMS temperature stabilizes at the set point. However, the $^3$He pot temperature does not reach the minimum temperature even after the MPMS temperature does to the set point. The magnetization measurement restarts sometimes even when the $^3$He pot temperature is not low enough, causing measurement errors as indicated by arrows in Fig. 6(a). We removed those data points and plot magnetization data in Fig. 6(b). The magnetization data are based on two sequential measurements.

The magnetization linearly decreases from zero field and start increasing after reaching minimum at $H_p$=70 Oe as shown more clearly in the inset. We estimated the $H_{C1}$=33 Oe at 0.53 K where the magnetization curve starts deviating from the perfect diamagnetic line based on the sample volume and $M = -H/4\pi$. The estimated $H_p$ and $H_{c1}$ matches well with the extrapolated values, $H_{C1}$=36 Oe, $H_p$=78 Oe, from the previous measurements down to 1.5 K[26] ensuring our measurement data quality. Further, from the comparison between the magnetization curve and the perfect diamagnetism line, the superconducting volume fraction is estimated to 100%. The magnetization curve shows discontinuous jump around 250 Oe as indicated by an arrow, followed by gradual increase up to 2000 Oe. This jump is reproducible in repeated measurements. The temperature dependence of the thermal conductivity of CeCoIn$_5$ also shows an anomalous broad peak in the similar field and temperature region; at $H$=200 Oe and $T \sim 0.5$ K[27]. However, we have no clear explanation about whether any common origin for both phenomena exists or just coincidence.

## 4. Conclusions

We have developed high-resolution magnetization measurement down to ~ 0.5 K with a miniature $^3$He cryostat and closed-cycle gas-handling system which is high-pressure-compatible. By condensing $^3$He gas at about ambient pressure, we successfully liquefy $^3$He gas at 3 K and in a relatively short condensing time. We confirm the measurement quality of this measurement setup with two types of magnetization measurement; the temperature dependence of the magnetization of aluminum to observe superconducting transition, the field dependence of the magnetization of the heavy fermion compound CeCoIn$_5$ to determine the lower critical field $H_{C1}$. This measurement system allows us to perform high-resolution magnetization measurement at very low temperatures down to 0.5 K relatively easy compared to the other magnetization measurement methods. Particularly, it has more advantage at very low field below 1000 Oe thanks to the high-resolution of a SQUID magnetometer.

## Acknowledgments

We thank Yuji Inagaki, Hiroyuki Tsujii, Koichi Matsumoto, and Koichi Nunomura for valuable discussions and technical supports. This work was partially supported by Grants-in-Aid for Scientific Research from the Japan Society for the Promotion of Science (no. 19K21845) and Shibuya science culture and sports foundation and Hitachi Metals-Materials Science Foundation.

**Figures and the Captions**

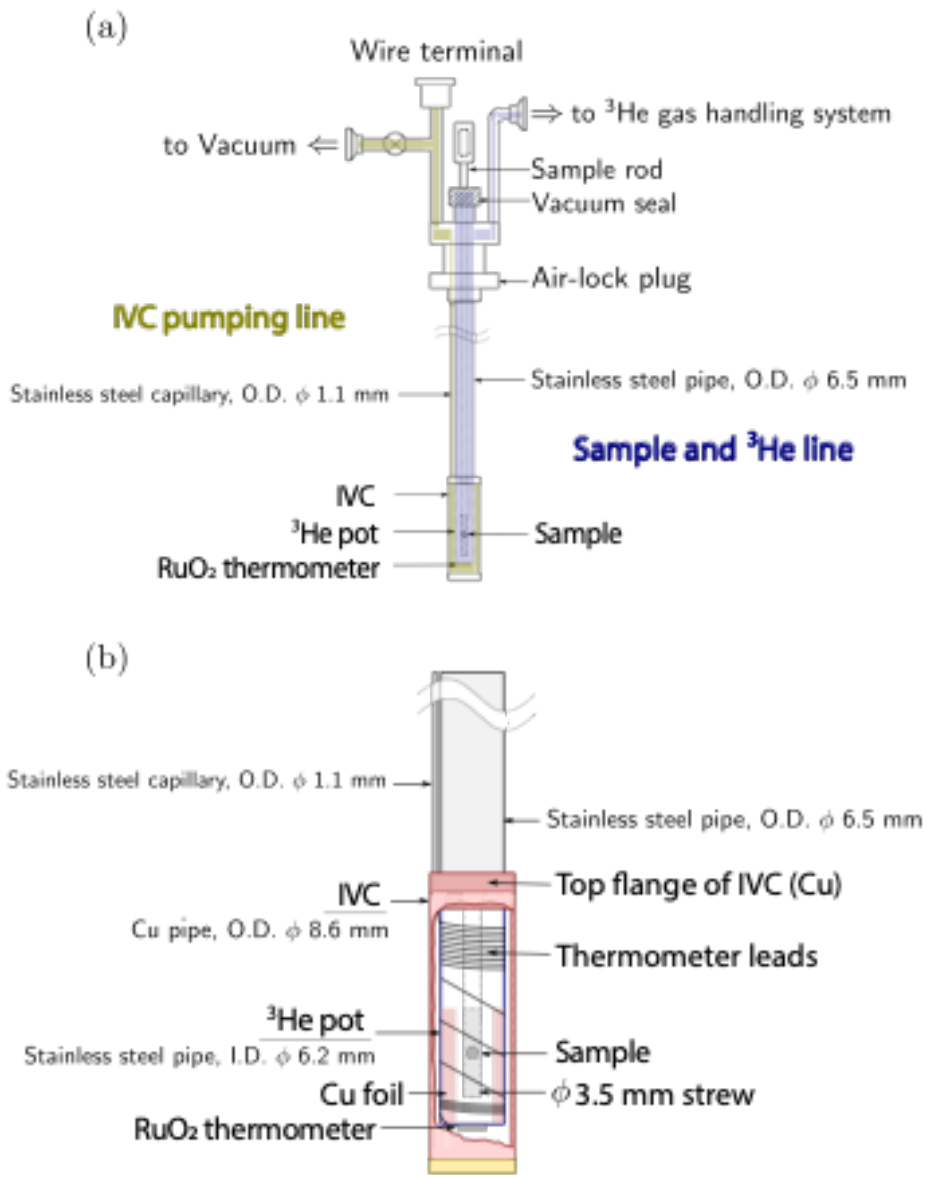

**Fig. 1.** Design of our miniature $^3$He cryostat. (a) Overview sketch of the whole part of the cryostat. (b) Focus view of the low-temperature area.

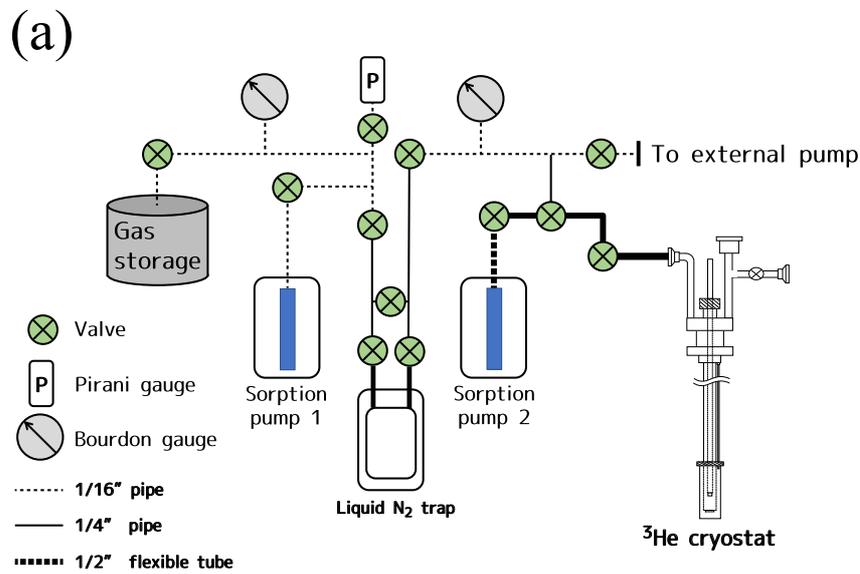

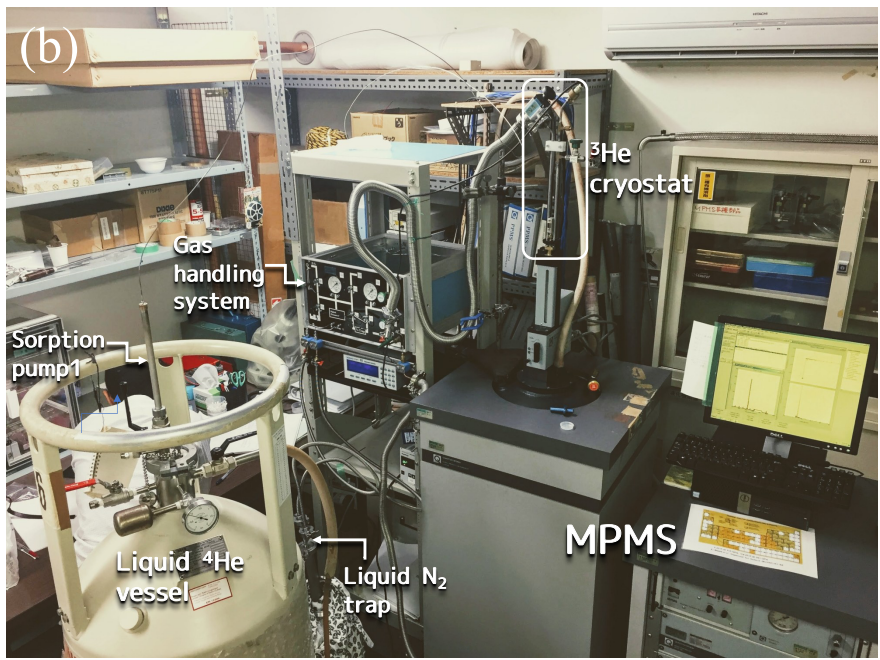

**Fig. 2.** (a) Schematic diagram of the gas handling system (GHS) and the miniature $^3$He cryostat. Sorption pumps 1 and 2 are dipped into a liquid $^4$He vessels, and the liquid N$_2$ trap is into a liquid nitrogen Dewar. (b) Photo of the measurement setup including the $^3$He cryostat attached to MPMS, the GHS, the liquid N$_2$ trap, and the sorption pump1 inserted to liquid $^4$He vessel

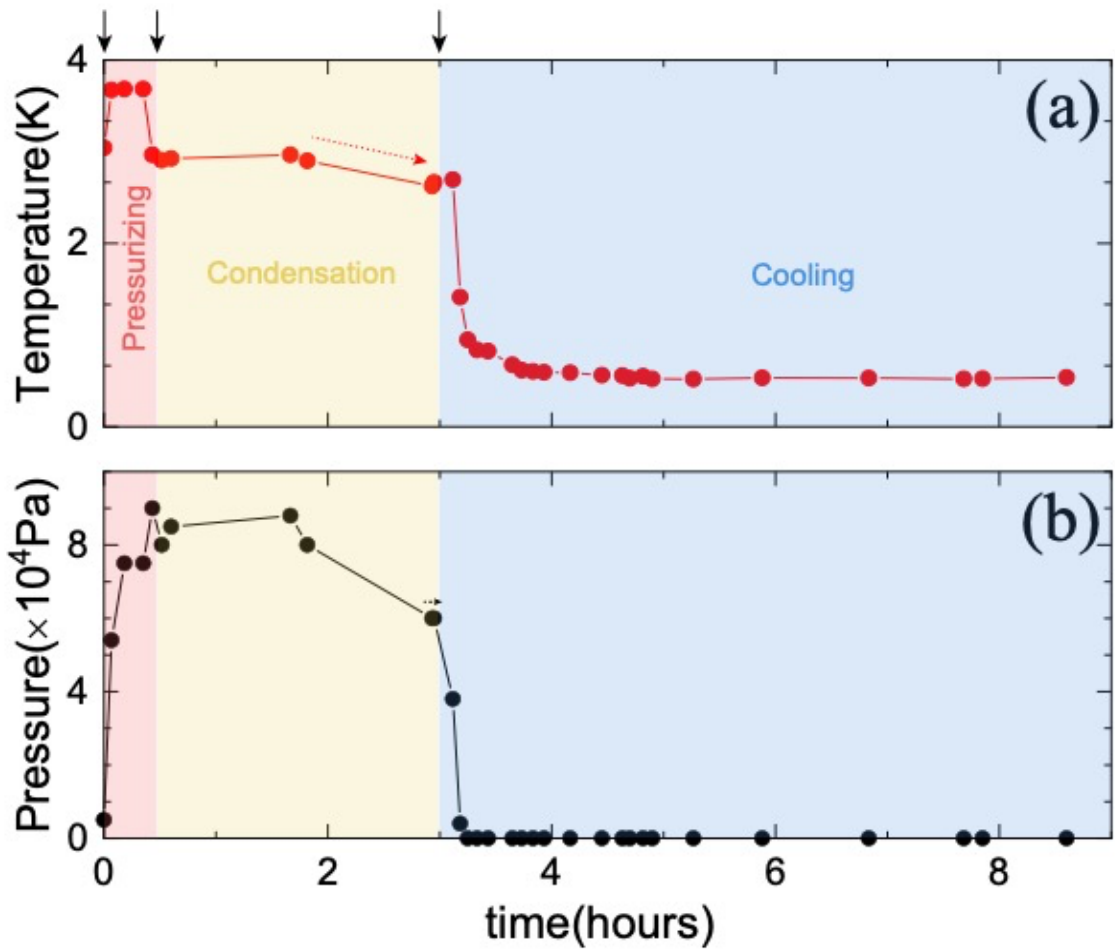

**Fig. 3.** Pressure variation of the GHS (a) and temperature variations of the $^3$He cryostat (b) with time. Solid arrows indicate the beginnings of $^3$He pressurization, $^3$He condensation, and $^3$He evacuation. A broken arrow in (a) indicates slow temperature decrease by lowering the MPMS temperature. A broken arrow in (b) indicates the stabilization of the $^3$He pressure in the GHS; the end of the $^3$He condensation at the given temperature.

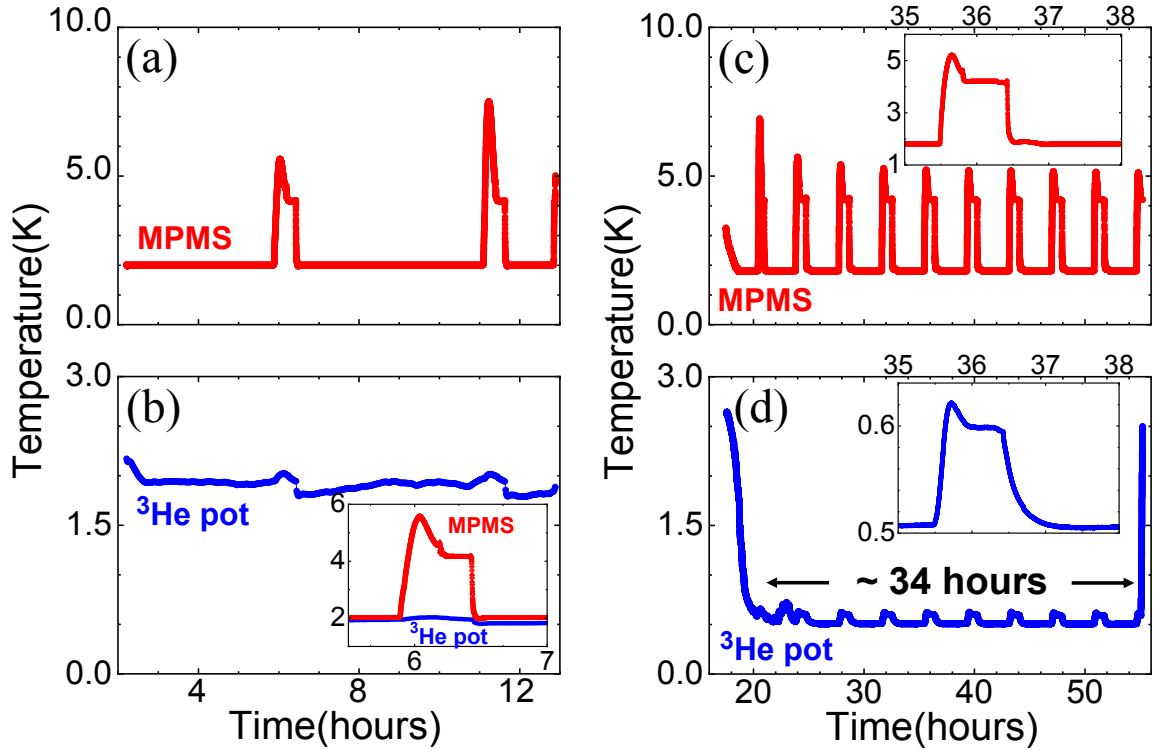

**Fig. 4.** (a) and (b): Temperature variations of MPMS and the $^3$He cryostat during $^3$He gas condensation. The inset in (b) shows comparison between changes in the MPMS and the $^3$He-pot temperatures, when the MPMS temperature rises. (c) and (d): Temperature variations of MPMS and $^3$He cryostat with time in the $^3$He cooling process. The insets in (c) and (d) show temperature variations in MPMS and $^3$He cryostat when MPMS temperature increases.

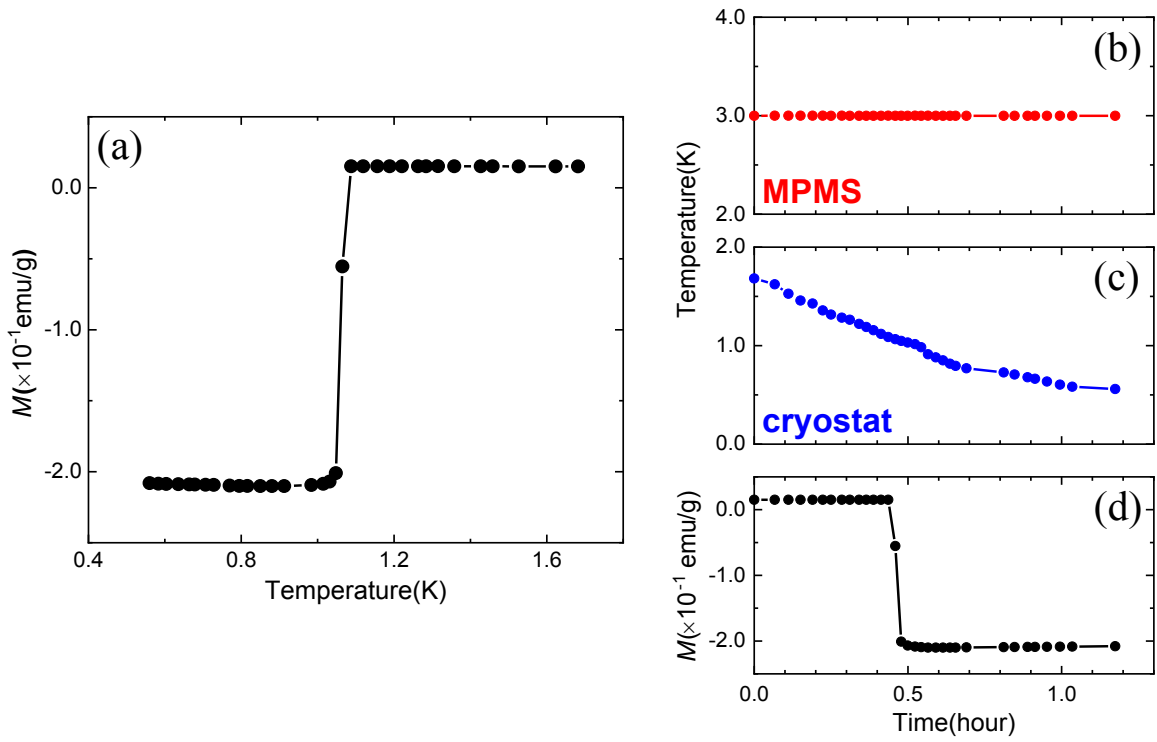

**Fig. 5.** (a) Temperature dependence of the magnetization of Al. Temperature variations of MPMS (b) and $^3$He-cryostat (c), and magnetization variation of Al (d) as a function of time during the magnetization measurement.

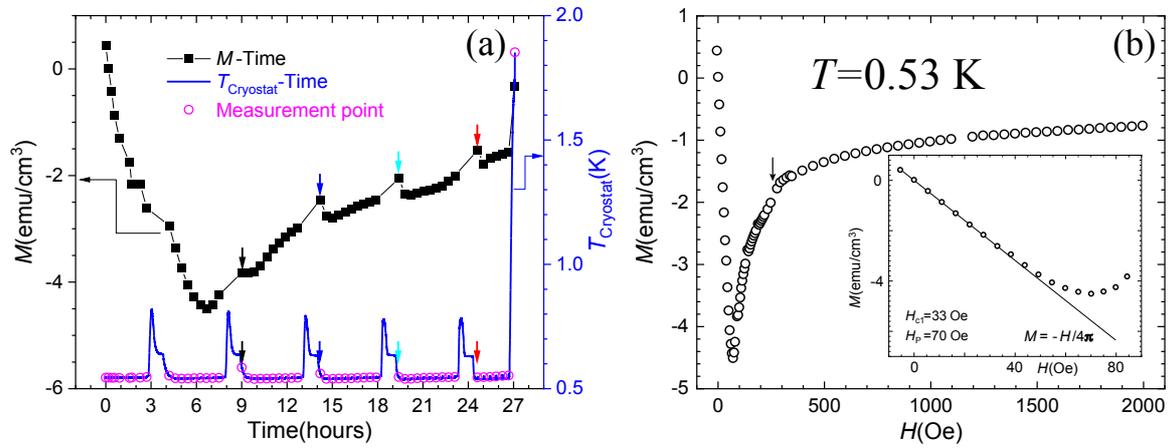

**Fig. 6.** (a) Magnetization of CeCoIn$_5$ and the $^3$He pot temperature as a function of time during magnetization measurement at 0.53 K. Open circles indicate the actual magnetization measurement points. Arrows indicate the errors in magnetization measurements due to higher measurement temperatures caused by periodic temperature rises of MPMS. (b) Magnetization curve of CeCoIn$_5$ at 0.53 K up to 2000 Oe. An arrow indicates a jump at 250 Oe. The inset shows comparison between the magnetization curve and calculated magnetization assuming perfect diamagnetism of the sample based on the sample volume.